# Generating Isolated Terawatt-Attosecond X-ray Pulses via a Chirped Laser Enhanced High-Gain Free-electron Laser

Zhen Wang, Chao Feng* and Zhentang Zhao

*Shanghai Institute of Applied Physics, Chinese Academy of Sciences, Shanghai 201800, China*

*Corresponding author: fengchao@sinap.ac.cn*

A feasible method is proposed to generate isolated attosecond terawatt x-ray radiation pulses in high-gain free-electron lasers. In the proposed scheme, a frequency chirped laser pulse is employed to generate a gradually-varied spacing current enhancement of the electron beam and a series of spatiotemporal shifters are applied between the undulator sections to amplify a chosen ultra-short radiation pulse from self-amplified spontaneous emission. Three-dimensional start-to-end simulations have been carried out and the calculation results demonstrated that 0.15 nm x-ray pulses with peak power over 1TW and duration of several tens of attoseconds could be achieved by using the proposed technique.

Free-electron laser (FEL) is such a device that utilize relativistic electron beams as gain medium to amplify the initial electromagnetic field. Unlike conventional lasers, the output photon energy range, temporal durations and peak brightness of FELs are theoretically unlimited, which makes FEL a unique and innovative approach for the realization of tunable, high-intensity, coherent x-ray source. To date, several x-ray FEL facilities have been constructed worldwide [1-5] and have already enabled observation and control of very fast phenomena at the atomic time scale, providing an ideal tool in various subjects such as femtochemistry, ultra-high resolution imaging, and the investigation of the dynamics in atomic and biological systems, etc [6]. Nowadays, most of the existing and the under-constructing x-ray FEL facilities [7-9] take advantage of the self-amplified spontaneous emission (SASE) scheme [10, 11], which can provide spatially coherent radiation pulse with high peak power at GW level and pulse duration of about 100 femtoseconds. One of the major goals for future XFELs is to further enhance the peak power and shorten the pulse duration to increase the resolution of x-ray diffraction imaging experiments and limit radiation damage on the samples simultaneously.

Several methods [12-19] have been developed in the last decade to manipulate the electron beam properties to generate ultra-short radiation pulses. A conceptually simple method is using low charge electron beam to reduce the electron bunch length as well as the radiation pulse length by about two orders of magnitude [12]. Alternately, one may adopt the emittance spoiler technique which employs a slotted foil in the central of the bunch compressor to select a small unspoiled part of the electron beam to lase [13]. Another concept, dubbed enhanced SASE (ESASE), take advantage of an external few-cycle laser to manipulate the electron beam longitudinal phase space to enhance the peak current in a short slice for generating attosecond radiation pulses [14]. The low charge operation and emittance spoiler techniques have been experimentally demonstrated at the linac coherent light source (LCLS) to generate hard x-ray pulses with a few femtosecond pulse durations [12, 15]. There is also a plan for developing the ESASE technique at LCLS in the near future [16]. However, none of these techniques hold the ability to enhance the output peak power beyond saturation, and the shortest output pulse length from ESASE is limited by the FEL slippage length, which is on the order of a few hundred attoseconds.

In order to overcome the FEL pulse lengthen from slippage effect, the mode-locking concept has been applied to SASE FELs for generating attosecond pulse trains. In the initial proposed scheme [17], a periodic variation in the electron beam energy or density was introduced by a high-power optical laser, then a mode-locked undulator, which consists of a series of undulator-chicane modules, is imposed to add precise delays between the electron beam and the radiation field, and thereby continually amplify the ultra-short x-ray pulse trains. Recently, a pulse-compression scheme [18] that combines the ESASE principle and the mode-locking technique has been proposed to achieve up to 100-fold increase in peak power of x-ray pulses, the x-ray beam can be delivered in about 50 attosecond at photon energies around 10 keV. However, in order to get isolated pulses, additional delay modules with optical mirrors to delay the x-ray with respect to the electron beam are needed in this scheme, and thus makes the layout relatively complex. Later on, an alternative method [19] based on the emittance spoiler technique was proposed to simplify the overall design. A multiple-slotted foil with uneven spacing of slots has been adopted to prevent the built up of multiple pulses and allows to amplify only one single short pulse by properly setting the strengths of delay sections. However, in this scheme, if a very narrow slit is used, the uncorrelated energy spread and betatron beam size will dominate the output slice length, and it will be very challenging to generate sub-femtosecond radiation pulses [15].

In this paper, we propose a novel and feasible scheme to generate isolated terawatt x-ray pulses with pulse durations of several tens of attoseconds. The proposed scheme is also based on the ESASE and superradiant principles [20, 21]. However, no x-ray delay sections are required, thus makes the proposed technique much simpler and can be easily applied to existing and future x-ray FELs. When compared with the emittance spoiler technique, the proposed technique holds the ability of achieving shorter output pulse durations and has the additional advantage of natural synchronization to external lasers, which is especially important for pump-probe experiments.

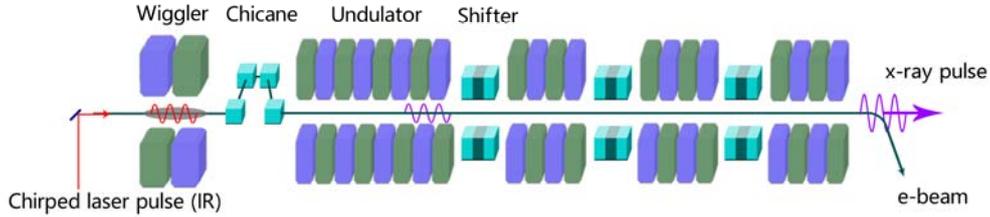

FIG. 1. Schematic layout of the proposed technique.

The schematic layout of the proposed technique is illustrated in Fig. 1, consisting of an ESASE section upstream of the undulator and some phase shifters inserted among the undulator sections. In the conventional ESASE scheme, an ultra-short laser pulse is employed to interact with the electron beam in a single period wiggler to introduce an energy modulation. Then the electron beam is sent through a dispersive magnetic chicane to convert the energy modulation into density modulation, which results in a periodic enhancement of the electron peak current. In order to get isolated radiation pulse, an ultra-short laser with no more than two optical cycles are required to minimize satellite spikes. This kind of few-cycle laser can be achieved through the chirped pulse amplification technique (CPA) [22], where a longitudinal stretched, frequency chirped laser pulse is amplified by the gain medium and sent through an optical element (compressor) with a suitable chromatic dispersion to recompress the beam to a bandwidth-limited pulse. The main challenge in pulse compression process lies in effective dispersion compensation and hence compression of the generated bandwidths to yield an isolated ultra-short optical pulse.

In the proposed scheme, instead of a few-cycle laser pulse, a frequency chirped laser is utilized to imprint a gradually-varied spacing current enhancement on the electron beam. This powerful laser pulse with significant and adjustable frequency chirp can be easily obtained by tuning the parameters of the compressor of CPA. The dispersion in the laser pulse do not need to be fully compensated, which would significantly simplify the laser system design for ESASE. After the density modulation, the electron beam is sent into a long undulator with a series of spatiotemporal shifters between the undulators to amplify a chosen ultra-short radiation pulse via superradiant gain process. In the first undulator section, the gradually-varied spacing current enhancement leads to an uneven spacing radiation pulse train. The first undulator should be relatively long to amplify the target radiation pulse with peak power that is high enough to suppress the shot noise in the following undulator section but still far from saturation to prevent degradation of the electron beam quality. The target radiation pulse is then shifted forward to the following current peak by the phase shifter while the other radiation pulses are shifted to the low-current part of the beam. The microbunching formed in the current peaks are smeared out by the phase shifter, preventing the growth of noisy spikes in the following undulators. The target radiation pulse reseeds the fresh peak in the following undulator section, leading to a continuing amplification of the radiation pulse. Finally, the target ultra-short pulse will be amplified to TW level, while the other pulses are suppressed to the noise level.

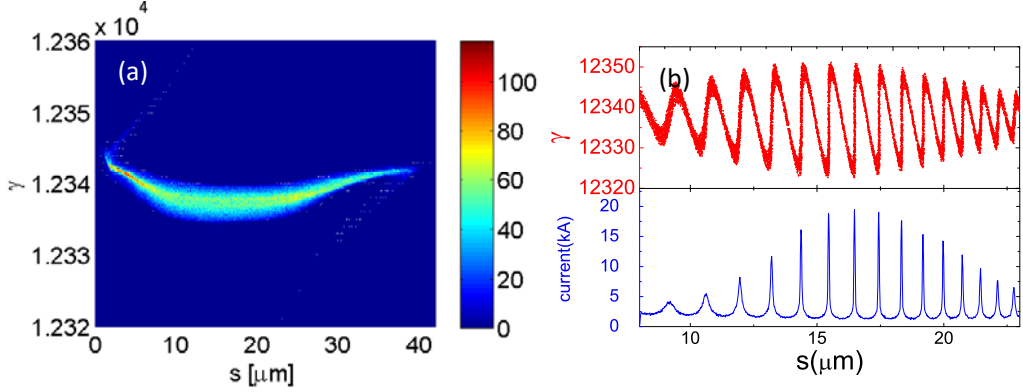

FIG. 2. (a) Longitudinal phase space at the end of the linac. (b) Longitudinal phase space and the corresponding current distributions at the end of the ESASE section. The bunch head is to the right.

To illustrate the physical mechanism and a possible application with realistic parameters, three-dimensional start-to-end simulations have been performed for the proposed technique. We take the layout and nominal parameters of the Shanghai hard x-ray FEL facility (HXFEL) [23], which is aiming at generating GW-level femtosecond radiation pulse based on SASE. The linac of HXFEL consists of an S-band photoinjector, a laser heater system, an X-band linearizer, two bunch compressors and three C-band main accelerator sections. At the exit of the linac, the electron beam can be accelerated to about 6.3 GeV with bunch charge of 250 pC and peak current of around 3 kA. The electron beam dynamics in the photoinjector was simulated with ASTRA [24] to take into account space-charge effects. ELEGANT [25] was then used for the simulation in the remainder of the linac. The longitudinal phase space of the electron beam at the end of the linac is shown in Fig. 2 (a), where one can find that a constant profile of beam energy, energy spread and emittance is maintained in an approximately 50 fs region in longitudinal. In this region, the normalized transverse emittances are around 0.4 mm mrad in both horizontal and vertical directions and the slice energy spread is about 600 keV. This part of electron beam is used for the FEL lasing. The ESASE and FEL gain processes was simulated by GENESIS [26]. A chirped laser pulse with central wavelength of 800nm, peak power of about 80 GW, beam waist of 200 μ m and pulse length of 40fs (FWHM) is injected to the ESASE modulator (a single period wiggler with period length of 20 cm). The bandwidth of the laser pulse is about 34%, equivalent to a 3.5 fsec transform-limit pulse after fully pulse compression.

The longitudinal phase space and current distributions of the electron beam at the exit of the ESASE section are shown in Fig. 2 (b), where a gradually-varied spacing current enhancement is achieved with peak current of about 20kA. The energy modulation amplitude induced by the laser pulse is about 10 times of the initial beam energy spread. The durations of the current spikes in the vicinity of beam central are on the scale of ~100 asec. This electron beam is then sent into to the following undulator sections to generate ultrashort radiation pulses at 1.5 Å. The simulation results for the x-ray radiation profile evolution in the undulator are illustrated in Fig. 3.

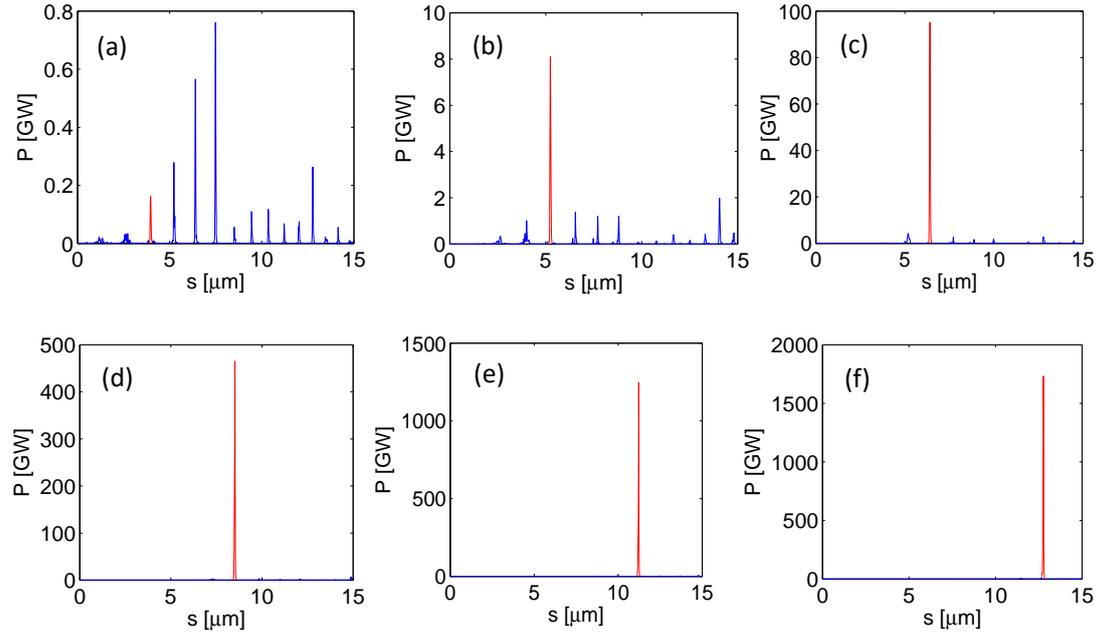

FIG. 3. XFEL radiation structure evolutions along the undulator beamline: (a) at the end of the 1st undulator section; (b) at the end of the 2nd undulator section; (c) at the end of the 3rd undulator section; (d) at the end of the 5th undulator section; (e) at the end of the 8th undulator section and (f) 10th at the end of the 10th undulator section.

The length of each undulator segment is about 2m (100 periods) with period length of 18 mm. The undulator segments in the 1st to 10th undulator sections are chosen to be 7, 4, 3, 2, 2, 2, 2, 2, 2, 2 to optimize the FEL power and improve the contrast of the target pulse against others. The first undulator section with 7 segments was used to produce an attosecond pulse train with hundred-megawatt-level peak power. The longitudinal profile of the radiation reflects the comblike structure of the current distribution, as shown in Fig. 3 (a). After that, phase shifters are applied to selectively amplify the target pulse (indicated in red) in the following undulator segments. The delay distances induced by these phase shifters are comparable to the initial laser wavelength. These small time delays can be easily realized by using small magnetic chicanes or short wigglers with the same type of the modulator.

In the second undulator section, the target radiation pulse was shifted forward and overlapped with the next current peak, leading to a continually power amplification. Beside the target radiation pulse, we found in the simulations that there are still some degree of overlap between other radiation pulses and current peaks due to the relatively long radiation pulse length (long slippage length) generated from the first undulator section. This results in the appearance of several satellite pulses in addition to the target pulse, as shown in Fig. 3 (b). However, the amplification of these satellite pulses dose not proceed as the slippage length becomes shorter and shorter in the following undulator sections. As shown in Fig. 3 (d), after 5 undulator sections, the peak power of the target radiation pulse is enhanced to about 500 GW, which is already two orders of magnitude higher than that of the satellite pulses. The contrast, which is defined by the percentage of the energy contained in the target pulse, is further improved in the following undulators. After 10 undulator sections, an isolated radiation pulse with pulse length of about 80 asec (FWHM) and the peak power of about 1.7 TW is produced, as shown in Fig. 3 (f). The contrast of the target radiation pulse is over 96%. The final output power can be further increased by utilizing more undulator sections.

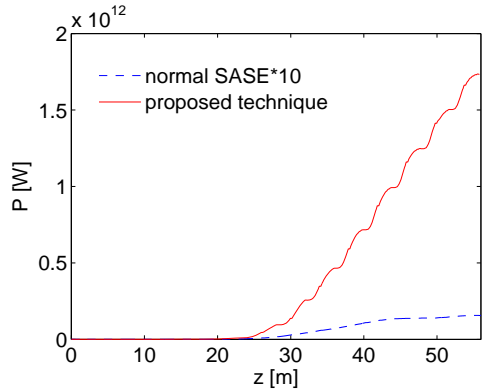

FIG. 4. FEL peak power gain curves for normal SASE (blue dashed line) and the proposed scheme (red solid line).

Fig. 4 shows the FEL peak power growths along the undulator distance. For comparison purpose, simulations were also performed for the normal SASE case (blue line) with the same electron beam and same undulators but without phase shifters. The normal SASE-FEL gets saturation at around 45 m of the undulator. The saturation power is about 13 GW. These simulation results demonstrated that the output peak power of SASE can be enhanced by about two orders of magnitudes by using the proposed technique (red line) with nearly same undulator length.

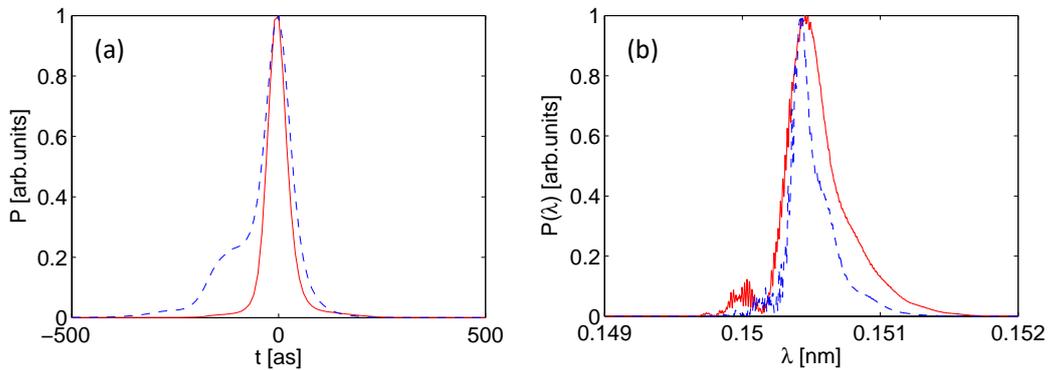

FIG. 5. Comparisons of the single shot radiation pulse (a) and corresponding spectra (b) from the proposed scheme with initial electron beam peak current of 3 kA (blue dashed line) and 1 kA (red solid line).

The final output pulse duration of the proposed scheme is determined by the widths of the current spikes and the slippage lengths in the last few undulator sections where the FEL power gain is quite large. For a given energy modulation amplitude, the width of the current spike is mainly determined by the initial uncorrelated energy spread of the electron beam [18], which can be tuned by changing the bunch compression factor of the linac. Here we performed simulations with the same parameters used above but an electron beam with peak current of 1 kA and initial energy spread of about 200 keV. The simulation results are summarized in Fig. 5. Although the durations of the current peaks have been reduced by about 3 times, the final output pulse duration is only reduced by about 1.5 times, from 80 asec to about 52 asec, as shown in Fig. 5 (a). The minimal output pulse duration is limited by the FEL slippage length which is about 100 asec in our case. One can further reduce the output pulse length by using shorter undulator sections, however, the FEL gain will be significantly degraded. The contrast of the target radiation pulse against background for this case is over 98%, which results in a quite clear output spectrum as shown in Fig. 5 (b). The FWHM bandwidth of the spectrum is about 0.2%, which is about 1.5 times wider than the Fourier transform limit.

In above simulations, the required density modulation is introduced by a chirped laser with very broad bandwidth, which may limit the applications of the proposed technique. We found later that the task can also be performed by utilizing commercial laser with central wavelength of 800 nm and spectral bandwidth of about 17% (equivalent to a 7 fsec transform-limit pulse). However, as the separation differences between adjacent current peaks become smaller, the lengths of the first few undulator sections should be reduced to prevent the power growth of satellite pulses. According to our simulations, the numbers of undulator segments in each undulator section are chosen to be 6, 3, 2, 2, 2, 2, 2, 2, 2, 2 to optimize the contrast. An isolated x-ray radiation pulse with pulse duration of around 80 asec, peak power of around 1 TW and contrast of about 85% can be generated at the exit of a 50 m long undulator.

In conclusion, a novel and easy-to-implement method has been proposed to significantly improve the output properties of a SASE-FEL. By making modifications of the laser system of ESASE and adding small phase shifters between undulator sections, terawatt-attosecond isolated radiation pulses could be produced. Three dimensional start-to-end simulations have been carried out to show the possible performance of the proposed technique with realistic parameters of a hard x-ray FEL facility, and the simulation results demonstrated that fully coherent hard x-ray radiation pulse with peak power at TW level and pulse duration less than 100 asec can be achieved in a ~50 m undulator. The output pulse length is mainly limited by the FEL slippage length in the undulator sections. One can further reduce the output pulse length and maintain the peak power by using better beam parameters, e.g. higher beam energy, smaller initial energy spread or lower transverse emittance, to enhance the FEL gain in a relatively short undulator. This kind of coherent x-ray light source is potentially useful where ultra-brightness and ultra-short pulses are required. It may open a new regime of ultrafast x-ray sciences, such as x-ray nonlinear optics and 3D atomic resolution imaging of single molecules [27, 28].

The authors would like to thank H.X. Deng, Q. Gu and D. Wang for helpful discussions and useful comments. This work is supported by the National Natural Science Foundation of China (11475250) and Youth Innovation Promotion Association CAS.

**Reference**
[1]  W.A. Ackermann *et al.*, Nature Photon. **1**, 336 (2007).
[2]  P. Emma *et al.*, Nature Photon. **4**, 641 (2010).
[3]  T. Ishikawa *et al.*, Nature Photon. **6**, 540 (2012).
[4]  E. Allaria *et al.*, Nature Photon. **7**, 913 (2013).

[5] C. Pellegrini, A. Marinelli, S. Reiche, Rev. Mod. Phys. **88**, 015006 (2016).
[6] C. Bostedt *et al.*, Rev. Mod. Phys. **88**, 015007 (2016).
[7] M. Altarelli *et al.*, The European X-ray free-electron laser, Technical Design Report, DESY (2007).
[8] R. Ganter, SwissFEL conceptual design report, PSI (2010).
[9] J.H. Han, H.S. Kang, I.S. Ko, in *Proceeding of the IPAC2012, New Orleans, USA, 2012,* p. 1735.
[10] A. Kondratenko and E. Saldin, Part. Accel. **10**, 207 (1980).
[11] R. Bonifacio, C. Pellegrini and L. M. Narducci, Opt. Commun. **50**, 373 (1984).
[12] Y. Ding *et al.*, Phys. Rev. Lett. **102**, 254801 (2009).
[13] P. Emma, K. Bane, M. Cornacchia, Z. Huang, H. Schlarb, G. Stupakov, and D. Walz, Phys. Rev. Lett. **92**, 074801 (2004).
[14] A. A. Zholents, Phys. Rev. ST Accel. Beams **8**, 040701 (2005).
[15] Y. Ding *et al.*, Phys. Rev. Lett. **109**, 254802 (2012).
[16] Z. Huang, in *Workshop on tailored soft x-ray pulses, PSI, Switzerland, 2016*.
[17] N. R. Thompson and B. W. J. McNeil, Phys. Rev. Lett. **100**, 203901 (2008).
[18] T. Tanaka, Phys. Rev. Lett. **110**, 084801 (2013)
[19] E. Prat and S. Reiche, Phys. Rev. Lett. **114**, 244801 (2015)
[20] R. Bonifacio, L. De Salvo Souza, P. Pierini, N. Piovella, Nucl. Instrum. Methods Phys. Res., Sect. A **296**, 358 (1990).
[21] R. Bonifacio, N. Piovella, and B. W. J. McNeil, Phys. Rev. A 44, R3441 (1991).
[22] P. Maine, D. Strickland, P. Bado, M. Pessot, G. Mourou, IEEE Journal of Quantum Electronics QE-**24**, 398 (1988).
[23] Z.T. Zhao *et al.*, In *Proceedings of the FEL2010, Malmo, Sweden, 2010,* p. 626.
[24] K. Floettmann, ASTRA User's Manual, available at http://www.desy.de/mpyflo/Astra_dokumentationS (1999).
[25] M. Borland, Phys. Rev. STAccel. Beams **4**, 070701 (2001).
[26] S. Reiche, Nucl. Instrum. Methods Phys. Res., Sect. A **429**, 243 (1999).
[27] M. Fuchs *et al.*, Nat. Phys. **11**, 964 (2015).
[28] A. Aquila *et al.*, Struct. Dyn. **2**, 041701 (2015).